# Time in the theory of relativity: inertial time, light clocks, and proper time

Mario Bacelar Valente

Abstract

In a way similar to classical mechanics where we have the concept of inertial time as expressed in the motions of bodies, in the (special) theory of relativity we can regard the inertial time as the only notion of time at play. The inertial time is expressed also in the propagation of light. This gives rise to a notion of clock – the light clock, which we can regard as a notion derived from the inertial time. The light clock can be seen as a solution of the theory, which complies with the requirement that a clock to be so must have a rate that is independent of its past history. Contrary to Einstein's view, we do not need the concept of "clock" as an independent concept. This implies, in particular, that we do not need to rely on the notions of atomic clock or atomic time in the theory of relativity.

1. Introduction

In classical mechanics time is inertial time, at least if we adopt Neumann and Lange's approach. The inertial time is expressed in the motions. We can determine the inertial time, e.g., by the distance covered by a free body: to equal distances correspond equal intervals of inertial time. In the theory of relativity[1] it is far from clear what notion or notions of time might be at play. Einstein considers that we need to rely on an independent concept – that of clocks (see, e.g., Einstein 1921a, 213; Giovanelli 2014). Also we need the assumption that the rates of these clocks are independent from their previous history, i.e. the paths they had in space-time, their accelerations, presence of electromagnetic fields, etc. (see, e.g., Einstein, letter to Hermann Weyl, 19 April 1918; Synge 1960, 106; Geroch 1972, 8). This leads to define the time at each position in an inertial reference frame in terms of these clocks, which to be consistent we should consider as being atoms (or atomic clocks), since these are the physical systems that experimentally warrant considering that the rates of clocks are independent of their past history (see, e.g., Einstein, letter to Hermann Weyl, 19 April 1918; Einstein 1921a, 214; Synge 1960, 105-6; Bacelar Valente 2016, 191-6). To define the coordinate time in an inertial reference frame, Einstein considers that it is necessary to synchronize these (atomic) clocks. For that, he adopts the propagation of light and establishes by definition that the "time" needed for light to travel between two points A and B is the same in both directions. In Einstein's approach, this "time" of propagation is measured in terms of the atomic time; there is no notion of time specifically associated with the light propagation itself.

      Here it is presented a view altogether different from this one. First of all, the "time" of propagation of light is, in a way similar to that of the motion of bodies in classical mechanics, an inertial time; the (inertial) propagation/motion of light expresses the inertial time. Also, in this paper, one rejects Einstein's view that "clock" is an independent concept necessary for the theory. From the point of view developed here, "clock" is a notion derived from the inertial time. From the propagation of light one defines light clocks, which can be seen as two bodies/mirrors with light bouncing between them. Contrary to Einstein, it is contended here that light clocks can be seen as solutions/models of the theory of relativity. These light clocks can "substitute" the atomic clocks to have a local time at each position in an inertial reference frame. All coordinate time measurements are then made in terms of light propagation.

---

[1] Here, by the theory of relativity we mean what is usually called the special theory of relativity. We will not consider Einstein's gravitation theory (the general theory of relativity). To be more precise, we adopt the formulation of special relativity as the theory of Minkowski space-time (see, e.g., Geroch 1972, 3-39; Friedman 1983, 125-149; Wald 1984, 59-66). In this theory, by light rays we mean electromagnetic waves in the so-called optical limit, such that the light rays "travel on null curves" (Geroch 1972, 37). Material bodies will be idealized as (point-like) particles, and we will be considering the space-time events corresponding to a particle, i.e. the worldline of the particle, which "describes completely [in space-time] the entire history (and future) of the particle" (Geroch 1972, 4).

To have a consistent notion of light clocks, it is necessary that they comply with what can be called the requirement (not the assumption) that these clocks have a rate that is independent of their past history. It turns out that in the theory of relativity (1/c of) the length of the timelike wordline of a material body is an invariant (i.e. it is independent of the adopted inertial reference frame), which has the dimension of time. Minkowski called it proper time and considered that it corresponded to the time "gone by" a body moving along the worldline (which if this body is a clock means that the proper time is equal to the time read off by the clock). If a clock reads off a time equal to (1/c of) the length of its worldline we can conclude that the clock has a rate that is independent of its past history.

As Fletcher (2013) showed, a sufficiently small light clock reads off a time which approximates to an arbitrary degree the Minkowski proper time. This means that it is a clock whose rate is in the limit independent of its past history. From this result, we can conclude that the adoption of light clocks (that are "solutions" of the theory) is consistent. Also, we do not need to resort to atomic clocks to have a physical system that reads off a time identical to the Minkowski proper time. Accordingly, the notion of light clock is a consistent notion of clock arising from the theory and based on the inertial time of light propagation. It seems that the only notion necessary in the theory is, in fact, that of inertial time. In particular, there is no need for atomic time or atomic clocks.

The paper is organized as follows. In section 2 we review some elements regarding the concept of inertial time as formulated in classical mechanics and set forward the equivalent notion in the theory of relativity. In section 3 the points mentioned above regarding light clocks and the Minkowski proper time are elaborated. It is made the case that there is in fact a notion of inertial time and that the theory can be formulated just in terms of it.

2. Inertial time in classical mechanics and in the theory of relativity

The notion of inertial time arose in the context of a criticism of Newton's notions of absolute space and absolute time (see, e.g., Lange 1885; see also Jammer 1993, 140-1; DiSalle 2009). The main driving force in this conceptual analysis was the clarification of the law of inertia as stated by Newton: "every body continues in its state of rest, or of uniform motion in a right line unless it is compelled to change that state by a force impressed upon it" (Jammer 1993, 123).

Addressing the law of inertia, Neumann tried to provide a meaningful notion of time by noticing that while saying that a body's motion is uniform without a previous notion of uniform time has no concrete meaning, one can use free bodies to define the uniform time. For just one free body, to stipulate that a free body travels equal distances in equal times is a convention without any experimental significance. However, if we consider two (or more) free bodies we can consider the question of what are the distances covered by other free bodies when the first covers equal distances (corresponding by convention to equal time intervals). In this case, we find that the other free bodies also cover equal distances. This "result" enables to restate the part of Newton's first law related to time: the motion of free bodies is such that to equal path distances of any free body corresponds equal path distances of any other free body. We can say that to successive path distances correspond successive time intervals, defining time by the motions of the free bodies (Neumann 1870; Barbour 1989, 655-6; Torretti 1983, 16-7). In this approach, we can consider one of the free bodies in motion as a clock in relation to which we describe the motion of the other free bodies (see, e.g., Barbour 1989, 655; Barbour 2007, 578).

Following Neumann's work, Lange defined a three-dimensional inertial reference frame in a way similar to the definition of the inertial time:

In exactly the same way as the one-dimensional inertial timescale could be defined through one single point left to itself [(a free body)], the three-dimensional inertial system can be defined through three points left to themselves. (Lange 1885, 253)

We consider three free bodies projected from a single point and moving in non-coplanar directions.

These three free bodies enable to define three Cartesian axes in relation to which they move rectilinearly, as all other free bodies (see, e.g., Barbour 1989, 658; Barbour 2007, 580; DiSalle 2009). As Barbour called the attention to:

Using any one of the three chosen reference bodies as a Neumann inertial clock, one can simultaneously verify that further bodies are moving uniformly as well as rectilinearly. (Barbour 2007, 579)

The motion of free bodies is not rectilinear in relation to an unobserved absolute space, but in relation to a spatial (inertial) reference frame determined by the free bodies themselves. We might rephrase the part of the law of inertia related to the rectilinearity of motion by making explicit that a free body moves rectilinearly in relation to an inertial reference frame.[2] In fact, Lange presented the law of inertia in terms of two definitions and two theorems in which the "construction" of the spatial reference frame from the motion of three free bodies is made explicit as well as the definition of the inertial time in terms of the motion of free bodies (Lange 1885, 253-4; see also Torretti 1983, 17).

What happens to the inertial time in the theory of relativity? To give an answer to this question let us first see how Einstein addresses the law of inertia (Newton's first law). It turns out that Einstein basically gave cursory definitions of inertial reference frames, in terms similar to that of classical mechanics. According to Einstein:

The inertial frame and time in classical mechanics are best defined together by a suitable formulation of the law of inertia: It is possible to determine time in such a way and to assign to the coordinate system such a state of motion (inertial frame) that, with reference to the latter, force-free material points undergo no acceleration. (Einstein 1923, 75)

In his view, "[special relativity] takes from earlier physics the assumption of the validity of Euclidean geometry for the possible positions of rigid bodies, the inertial frame, and the law of inertia" (Einstein 1923, 76). In Einstein's view, in the context of the theory of relativity, "it is possible to choose [an inertial reference frame] that is in such a state of motion that every freely moving material point moves rectilinearly and uniformly relative to it" (Einstein 1915, 249). As it is, this "definition" of inertial reference frame seems to be inconsistent in the context of the theory of relativity (that it is incomplete was noticed by Torretti (1983, 51)). One is defining the inertial reference frame using the law of inertia. However, the law of inertia, in its standard formulation, seems to require first a definition of distant simultaneity in the inertial reference frame. In terms of Einstein's approach, to say that a free body travels equal distances in equal times presupposes the synchronization of the clocks of the reference frame that will measure the time gone by the free body when moving rectilinearly. But to synchronize the clocks we first consider them to be part of the inertial reference frame (see, e.g, Einstein 1905, 141-2; Einstein 1907b, 255-7; Einstein 1910, 125-8). It seems that we would have a circularity in this definition. Torretti's approach to overcoming the incompleteness of Einstein's approach (which in our view is inconsistent due to its circularity) also solves the circularity problem. In Torretti's view, the incompleteness can be avoided, following Einstein's own views, by defining the (spatial) inertial reference frame in relation to the rectilinear motion of free bodies and the rectilinear propagation of light rays. According to Torretti:

If an inertial and a non-inertial frame move past each other with uniform acceleration, a light-ray emitted through empty space in a direction normal to the mutual acceleration of the frame, describes a straight line in the inertial, a curved line in the other. The rectilinear propagation of light in vacuo

---

2 We will not address here what notion of "rectilinearity" is at play in the law of inertia, neither the related issue of the Euclidean character of a spatial reference frame in inertial motion (for some ideas related to these issues see, e.g., Bacelar Valente 2014, 3-6; Pfister and King 2015, 14-9).

provides therefore an additional criterion for the identification of inertial frames, which, be it noted, does not presuppose a definition of time. (Torretti 1983, 51)

According to Torretti, an inertial reference frame F is one in relation to which: a) Three free particles projected non-collinearly from a point in F describe straight lines (which corresponds to the first part of Lange's reformulation of the law of inertia, which only addresses the rectilinearity of the inertial motion, not its temporal uniformity); b) A light ray transmitted through empty space in any direction, from a point in F, describes a straight line (Torretti 1983, 51). This avoids, for now, any reference to the uniformity of time, as it is made in the law of inertia. The "inertial motion" is just characterized, e.g., in terms of the rectilinear motion of free bodies in relation to a (spatial) inertial reference frame (without any reference to the uniformity of time).

At this point, we can adopt an approach similar to that of Neumann and Lange. In this case, the inertial time is expressed by the propagation of light. We can see the so-called postulate of the constancy of the two-way speed of light as our definition of inertial time in a way similar to the definition in classical mechanics in terms of the motion of free bodies.[3] If light is sent from a body at rest in our (spatial) inertial reference frame and reflected back at a distance d from the body (as measured, e.g., using unit-measuring rods), the inertial time associated with the propagation of light is 2d/c, where c is the constant two-way speed of light.[4] In what follows we will see that this notion of time is the only notion necessary in the theory of relativity and gives rise to a consistent notion of light clock.

3. Light clocks and the Minkowski proper time

Adopting Einstein's views it might seem that it is not possible to adopt the notion of inertial time in the theory of relativity. According to Einstein, we need the notion of transportable identical clocks whose rates are independent of their past history to justify that the line element of the Minkowski space-time is invariant:

The quantity [ds] which is directly measurable by our unit measuring-rods and clocks … is therefore a uniquely determinate invariant for two neighboring events (points in the four-dimensional continuum), provided that we use measuring-rods that are equal to each other when brought together and superimposed, and clocks whose rates are the same when they are brought together. In this the physical assumption is essential that the relative lengths of two measuring-rods and the relative rates of two clocks are independent, in principle, of their previous history. (Einstein 1922, 323)

In a letter to Weyl, Einstein made clearer why is this assumption necessary in relation to the invariant ds:

Imagine two clocks running equally fast [(i.e. with the same rate)] at rest relative to each other. If they are separated from each other, moved in any way you like and then brought together again, they will again run equally (fast), i.e., their relative rates do not depend on their prehistories.

I imagine two points $P_1$ & $P_2$ that can be connected by a timelike line. The timelike elements $ds_1$ and $ds_2$ linked to $P_1$ & $P_2$ can then be connected by a number of timelike lines upon which they are

---

3 Obviously, the motions of bodies also express the inertial time. However, even for the simplest case of free bodies, the law of inertia in its standard formulation might be conventional (see, e.g., Bacelar Valente 2017). Adopting the light propagation and the light postulate to define the inertial time we avoid any eventual problem arising from a (possibly) conventional definition of inertial time.

4 Here, we are simplifying the presentation and, implicitly, refer to the value of the inertial time of light propagation (corresponding to one "cycle" of light bouncing between two bodies in relative rest in inertial motion) in terms of an independently measured distance d and an independently measured two-way speed of light c (e.g. by using unit-measuring rods and atomic clocks). We mention below that we do not need to resort to independently measured d and c to have a notion of inertial time, a measure of distance d, and a value of c.

lying. Clocks traveling along these lines give a fixed relation $ds_1 : ds_2$ independent of which connecting line is chosen.–if the relation between ds and the measuring-rod and clock measurements is dropped, the theory of rel. loses its empirical basis altogether. (Einstein, letter to Hermann Weyl, 15 April 1918)

According to Einstein, this assumption of the theory is warranted by experiment:

If two ideal clocks are going at the same rate at any time and at any place (being then in immediate proximity to each other), they will always go at the same rate, no matter where and when they are again compared with each other at one place. If this [assumption was] not valid for natural clocks, the proper frequencies for the separate atoms of the same chemical element would not be in such exact agreement as experience demonstrates. The existence of sharp spectral lines is a convincing experimental proof of [this assumption]. (Einstein 1921a, 213-4)[5]

This means that the rate of the clock is independent of the space-time path of transfer (Einstein 1921b, 225), which is equivalent to saying that the rate of the clocks is not affected by their acceleration or, e.g., the presence of electromagnetic fields (see, e.g., Einstein, letter to Walter Dällenbach, after 15 June 1918; Geroch 1972, 8).

    The existence of identical atoms (natural clocks) whose "proper frequencies" are independent of their past history justifies the adoption of the concept of transportable identical clocks. Einstein is, *de facto*, using (previous to its establishment) the notion of atomic time in his approach to time in the theory of relativity (see, e.g., Bacelar Valente 2016, 191-6). More than this, Einstein calls the attention to the fact that the atom (clock) is not described by the theory. This is the case since at the present time its best description is given by quantum mechanics. An atom (a clock) is not described as a "solution" of special relativity (or general relativity, for that matter).[6] According to Einstein:

[The concepts of rod and clock] must still be employed as independent concepts; for we are still far from possessing such certain knowledge of the theoretical principles of atomic structure as to be able to construct solid bodies and clocks theoretically from elementary concepts. (Einstein 1921a, 213; see also Giovanelli 2014)[7]

---

5 Einstein had noticed early that atoms can be seen as clocks (see, e.g., Einstein 1907a, 232; Einstein 1907b, 263; Einstein 1910, 134). In his view, "Since the oscillatory phenomena that produce a spectral line must be viewed as intra-atomic phenomena whose frequencies are uniquely determined by the nature of the ions [(atoms)], we can use these ions [(atoms)] as clocks" (Einstein 1910, 124-5). To be rigorous atoms are not yet clocks. We might consider that the reference to atoms as clocks points to the possibility of considering atomic clocks.

6 Since, as is defended in this paper, we can implement a notion of inertial time in the theory of relativity, we might equally consider that quantum mechanics or quantum electrodynamics implement this notion due to their reference to inertial coordinate systems (see, e.g., Dickson 2004, 201). In this way, we might be tempted to say that, contrary to Einstein's views, we do have a description of clocks in terms of an "underlying" inertial time using quantum theory. This does not have to bear on the views developed in this paper on two accounts: a) the description of atomic clocks would still be made outside the theory of relativity, b) while it might rely on a notion of inertial time (associated with the Galilean or Minkowskian space-time), we can, following Einstein, regard the description of clocks in terms of a "solution" of quantum theory as not corresponding to a hypothetical solution made within a single relativistic theory that provides "from within" a description of matter. In this way, in this paper it will not be considered that there is a relativistic description of atoms (in the sense given by Einstein). However, we will move away from Einstein's view, in relation to physical systems that are (in our view) described in the theory of relativity, which we can consider as clocks satisfying the theory's requirement that the clock's rate must be independent of the clock's past history.

7 While both Geroch (1972) and Synge (1960) present special and general relativity along the lines of Einstein's ideas, i.e. in terms of atomic clocks and the independence from past history assumption, they do not mention explicitly the idea that "clock" is an independent concept. Synge, in particular, mentions the assumption of "the existence of standard clocks" (Synge 1960, 105), which he regards as atoms (Synge 1960, 106). According to him, the time reading of standard/atomic clocks is "the only time of basic importance in relativity" (Synge 1960, 105), and "since all clocks henceforth considered are standard clocks, we shall drop the adjective and call them simply clocks" (Synge 1960, 107). In Synge's approach, "time is the only basic measure. Length (or distance), in so far as it is necessary or desirable to introduce it, is strictly a derived concept" (Synge 1960, 108).

Here we will make the case that, contrary to Einstein's view, there is, in fact, a notion of a clock, derived from the inertial propagation of light, intrinsic to the theory, which enables to consider that the time of the theory of relativity is the inertial time. This means that we can dispense with the notion of an (atomic) clock as an independent concept of the theory. For a question of consistency, it will be necessary that this intrinsic notion of a clock corresponds to the idea of a clock whose rate is independent of its past history.

But what do we mean by a clock? Like in the case of classical mechanics, a clock can be seen as a physical system that, in Barbour's words, "must 'lock onto' or 'tap' processes directly and exclusively governed by the local inertial frame of reference" (Barbour 2007, 581). This is due to the fact that "the inertial frame of reference and distance traversed in it are (in mechanics at least) always the ultimate source of a scientifically meaningful definition of time" (Barbour 2007, 581). Also, as Barbour mentions elsewhere, "it is not a clock that we must define but clocks and the correlations between them as expressed in the marching-in-step criterion" (Barbour 2009, 7). The theory of relativity enables a definition of a type of identical clocks based directly on the propagation of light – the light clocks.

We can consider two related notions of light clock, a pre-Minkowskian one, as mentioned, e.g., by Einstein (1913, 207), and a Minkowskian space-time notion as set forward by Marzke and Wheeler (1964). According to Einstein:

If two mirrors are placed at the ends of a natural length $l_0$ so that they face each other, and a vacuum light ray is made to pass back and forth between them, then this system represents a clock (light clock). (Einstein 1913, 207)

This idealized light clock can be further "abstracted" in terms of worldlines of the Minkowski space-time. According to Marzke and Wheeler:

Having to particles moving along parallel world lines, we can let a pulse of light be reflected back and forth between them. In this way we define a geodesic clock. It may be said to "tick" each time the light pulse arrives back at the object number one. (Marzke and Wheeler 1964, 53)[8]

In this case, the bodies can be described in a very general way in terms of timelike worldlines of the Minkowski space-time, while the light rays are described simply in terms of null worldlines (Marzke and Wheeler 1964; see also Ohanian 1976, 192-5, Fletcher 2013). A light clock in inertial motion consists in the straight timelike wordlines of two free particles (that are parallel to each other, keeping the same distance) with null geodesics of light rays "bouncing" between them (see, e.g., Ohanian 1976, 193). We do not need a standard of distance (e.g. a unit-measuring rod) to construct a parallel to a chosen straight worldline. As mentioned by Ohanian, "there is a way to construct parallels which does not involve any length measurements" (Ohanian 1976, 193). This is the Marzke-Wheeler construction, which shows how to construct parallel timelike worldlines only given a (spatial) inertial reference frame (Marzke and Wheeler 1964, 50-2; see also Ohanian 1976, 193-5).[9] According to Ohanian, "we can now take as the unit of time the interval between two "ticks" of the [light clock]. We can take as the unit of length the distance that light travels in one unit of time" (Ohanian 1976, 195). In this way, we also settle the value of the two-way speed of light c.[10]

---

8 We see that Marzke and Wheeler (1964) actually named the clock "geodesic clock", not "light clock". Ohanian (1976) adopted the name "geometrodynamic clock". The geodesic, geometrodynamic, or light clock is independent of the structure of matter and enables to measure space-time intervals (Marzke and Wheeler 1964, 53-8; Ohanian 1976, 192-200).

9 As Marzke and Wheeler call the attention to, their construction of parallel worldlines "depends upon the existence of an inertial reference system– a system in which the world lines of all light rays and all free particles appear straight" (Marzke and Wheeler 1964, 52).

10 To simplify the discussion we will consider that the units of time, length, and velocity (determined in terms of a particular light clock in inertial motion adopted as our standard) were chosen so that one "tick" of any other light clock,

Contrary to Einstein's view (adopted also by, e.g., Synge (1960, 106) and Geroch (1972, 8)), the independence from past history does not have to be seen as an independent assumption regarding a physical system not described by the theory. It is a necessary consequence of the theory: clocks to be so must "'lock onto' and reflect the inertial spatiotemporal framework" (Barbour 2007, 587), which in the context of special relativity we can rephrase and say that their time readings must correspond to an invariant ds (i.e. they cannot get "unlocked"). This implies, as noticed by Einstein, that the rates of the clocks must be independent of their past history. This means that for the time being we must consider the adoption of light clocks as provisional. In fact, we cannot even say that the physical systems in question are truly clocks. For that we must show that when adopting these physical systems (that are described by the theory) as clocks it turns out that their rates are independent of their past history.

But, previous to addressing this issue, we need first to ask if we can actually regard these physical systems (that we expect will have the physical behavior of clocks) as solutions of the theory? According to Einstein's views, this would not be the case. To Einstein, e.g. a light clock could not be taken to represent the concept of clock as a "solution" of the theory, in the sense given by him. While Einstein mentioned light clocks and other type of "inertial clocks" (see, e.g., Einstein 1911, 344; Einstein 1913, 207), which at first sight we might take to be described by special relativity, the theory does not provide a relativistic theory of matter as Einstein expect would occur "at a later stage of the theory" (Einstein 1949, 61).[11] We might speculate that from Einstein's point of view we might consider that instead of a dynamical description of light clocks, what we have are "just" simplified "representations" consisting in timelike worldlines taken to represent mirrors and null worldlines representing light rays. We cannot agree with this (hypothetical) "view" by Einstein.

It is a fact that a light clock is only described "kinematically" in terms of the timelike worldlines of particles and null worldlines of light rays, but these constitute a very particular physical system with very specific physical characteristics. In particular the construction of the [light clock] guarantees that the time measured by this clock coincides with the time variable t that appears in the equations of motion of a particles, in the Maxwell equations, in the Lorentz transformation equation, etc. We may express this by saying that the [light clock] time scale agrees with the inertial time scale. (Ohanian 1976, 195)[12]

Like Einstein, we might prefer to have a solution of differential equations to which we might relate concepts we call "dynamical". However, the physical system proposed by Marzke and Wheeler (1964) is a very specific physical system (constituted by two particles with light rays bouncing between them), which has very particular kinematic/geometrical properties (making it meaningful to name it): a) in the case of an inertial motion the particles' geodesics are parallel (keeping a constant distance); b) there is light bouncing between the particles; c) geometrically, successive bounces of light rays are identical (we have so to speak a sequence of identical zig-zag tracks of light; see, e.g., Ohanian 1976, 193); d) the time measured by the light clock is identical to the inertial time (taken to have been defined by other means).

At this point, we have a type of physical systems – "light clocks" – which are solutions of special relativity and which we expect will have rates that are independent of their past history so that we can say we have a consistent implementation in the theory of the notion of a clock. Right now we also only have implemented the notion of a spatial inertial reference frame. We need to complete it by establishing its coordinate time. For that we can adopt Einstein's approach and consider that we have identical "light clocks" spread on the spatial inertial reference frame which

---

moving inertially, is equal to 2d/c seconds, where d is the distance between the bodies/mirrors (abstracted as particles) and c is the two-way speed of light.
11 We must recall that Einstein expected that his relativistic theories might be "completed" in the future in the form of a unification theory that besides unifying gravitation and electromagnetism would give a field theoretical description of matter (see, e.g., Goenner 2004).
12 In this work we go a "step further" and define the inertial time scale in terms of the inertial propagation of light by using a light clock.

must be synchronized to have a coordinate time. Let us consider two identical clocks located at points A and B of a spatial inertial reference frame; according to Einstein:

It is not possible to compare the time of an event at A with one at B without a further stipulation … the latter can now be determined by establishing by *definition* that the "time" needed for the light to travel from A to B is equal to the "time" it needs to travel from B to A. For, suppose a ray of light leaves from A toward B at "A-time" $t_A$, is reflected from B toward A at "B-time" $t_B$, and arrives back at A at "A-time" $t_A'$. The two clocks are synchronous by definition if $t_B - t_A = t_A' - t_B$. (Einstein 1905, 142)

As Einstein expressed with clarity elsewhere, after the synchronization of all the clocks of the inertial reference frame we have a meaningful notion of coordinate time:

The aggregate of the reading of all clocks synchronized according to the above … we call the [coordinate] time belonging to the coordinate system used. (Einstein 1907b, 256)

We define the time coordinate of an event taking place at an arbitrary point of [the coordinate system] (point event) as the simultaneous reading of the clock set up at this point and regulated according to the given procedure. Two point events (occurring at different points) are simultaneous if their time coordinates are equal. (Einstein 1912-1914, 30)

We can see the "A-time" and "B-time" as given, in Einstein's approach, in terms of the atomic time. Also, the "time" taken by light to propagate between the two points A and B is measured in terms of the atomic time of the atomic clocks at A and B. However we can dispense with the notion of atomic clock and reframe this approach in terms of light clocks and the inertial time (expressed in the propagation of light). Let us consider a "light clock" constituted by two mirrors/bodies (at a position A and a close-by position) with light bouncing between them, at rest in the inertial reference frame. Every time the light bounces back at A it corresponds to a "tick" of the clock. We consider an equivalent system located at a position B. We synchronize these "light clocks" by exchanging light. There is a huge difference in relation to Einstein's original approach. We do not have a clock at A with an "A-time" and a clock at B with a "B-time", plus the "time" of the propagation of light. We only have the "time" (inertial time) associated with the propagation of light. The "A-time" and the "B-time" are derived notions, as it is the case with the notion of "light clocks", since the "A-time" and the "B-time" are measures of inertial time made by "light clocks" (in terms of the propagation of light). What we have is the inertial time as given by the light propagation.[13]

As mentioned previously, Einstein regarded the assumption of the clock's rate independence from its past history as necessary for the theory to be meaningful – in relation to the invariance of the line element ds. For a formulation of the notion of time in the theory of relativity totally in terms of the inertial time, it is necessary that the derived notion of "light clocks" satisfies this requirement. If this is the case, then we might say that in a self-consistent way the adoption of light clocks as clocks that measure inertial time (and coordinate time) at any location in the inertial reference frame, does not lead to any contradiction. To show that this is the case we will take into account the Minkowski proper time. If we consider the integral of (1/c of) the line element ds along a timelike worldline (which corresponds to a worldline of a material system), it is invariant, i.e. independent of the adopted coordinate system, and, as we can check immediately, it has the dimension of time (e.g. the second). Minkowski called this invariant the proper time.[14] It is usually regarded as the time gone by an (ideal) clock along the worldline (see, e.g., Bacelar Valente 2016).This brings issues like the so-called clock hypothesis, which is regarded by some as another assumption necessary to consider that a physical system behaves along a "reasonable" world line as an ideal clock (taken to be defined as a clock that gives/reads values corresponding to the proper time).[15]

---

[13] We must notice that with this approach in terms of inertial time the so-called second postulate comes before the first postulate. It is only after having completed inertial reference frames with time coordinates that it makes sense to postulate that different inertial reference frames are equivalent for the description of physical phenomena.

[14] The dimension of (1/c of) the length of a timelike worldline (the Minkowski proper time) is [1/c ∫ds] = [∫sqrt(1 − $v^2/c^2$) dt] = second.

The relevance of the Minkowski proper time for us is that a physical system that along a timelike wordline gives/reads a time according to the Minkowski proper time is a physical system that is independent of its past history. Looking into Minkowski's expression for the proper time $\tau = \int ds/c = \int sqrt(1 - \upsilon(t)^2/c^2)dt$, we notice that it is, as Brown puts it, "a sum ... of 'straight' infinitesimal elements [ds/c]" (Brown 2005, 95). For each infinitesimal straight (invariant) element $d\tau = ds/c$, the only difference with dt is the time dilation factor $d\tau = sqrt(1 - \upsilon(t)^2/c^2)dt$ that depends only on the instantaneous velocity $\upsilon(t)$. The (instantaneous) rate of the clock does not depend on how the clock has been accelerated or its path in space-time, i.e. it does not depend on its past history. In this way, if it turns out that the light clock gives a time along a timelike worldline whose value is identical to (1/c of) the length of the worldline, then it is independent of its past history. There is, in this case, no inconsistency in the early adoption of the notion of light clock, which must, implicitly, be taken to be independent of its past history to meaningfully relate it to the invariant line element, previous to actually addressing if the light clock is, in fact, independent of its past history. It is, as we have called it, a self-consistent implementation of a notion of a clock.

That this is the case can be concluded from Fletcher's proof of a theorem that implies that the time reading of a sufficiently small light clock can approximate to an arbitrary degree the length of a closed $C^{(2)}$ timelike worldline segment (i.e. the Minkowski proper time).[16] According to Fletcher:

One can then state the theorem in words as follows. Given a closed segment of a timelike curve and any $\varepsilon_A$, $\varepsilon_R > 0$, there is a sufficiently small and unvarying light clock that measures the [length of] that segment within an accuracy of $\varepsilon_A$ and ticks with no more than $\varepsilon_R$ variation in regularity. (Fletcher 2013, 1382)

---

15 Different authors give different meanings to the words "proper time", "clock hypothesis", and "ideal clock". To Brown "proper time" is the time read off by a clock (Brown 2005, 29 and 115). To establish the equality of the clock's time reading with the "integration of the metric along an *arbitrary* time-like curve" (Brown 2005, 9), i.e. to what Minkowski called the proper time, we need to resort to an assumption – the clock hypothesis, or as Brown calls it more recently the clock condition (Brown and Read 2016, 14-5). Accordingly, the clock hypothesis "is the claim that when a clock is accelerating, the effect of motion on the rate of the clock is no more than that associated with its instantaneous velocity – the acceleration adds nothing. This allows for the identification of [1/c $\int ds$] with the proper time [(as the time read off by the clock)]" (Brown 2005, 9). To Brown, the clock hypothesis is "not a consequence of Einstein's 1905 postulates" (Brown and Pooley 2001, 264). Also, to Brown an ideal clock is a clock that when in inertial motion reads off inertial time (Brown and Read 2016, 14). We need then, in his account, to assume the clock hypothesis/condition to consider that the time reading of an ideal clock in accelerated motion is equal 1/c $\int ds$ (which Minkowski called the proper time). To Arthur "proper time" is, as defined by Minkowski, 1/c $\int ds$. To Arthur, proper time is a physical quantity predicted by the theory of relativity, since it is invariant, i.e. independent of the particular inertial reference frame adopted (Arthur 2010, 177). Also, there is a notion that is predicted by the theory; that of an ideal clock (Arthur 2010, 159): an ideal clock is one that reads proper time (Arthur 2010, 166). Regarding the clock hypothesis, it is "not needed as an independent postulate in [the theory of relativity]. Insofar as it can be regarded as stating the criterion for an ideal clock in [the theory of relativity], it is already implicit in that theory in the invariance of proper time" (Arthur 2010, 177). Accordingly, "the argument that many real clocks will fail to satisfy the clock hypothesis is just the claim that many processes fail to qualify as ideal clocks" (Arthur 2010, 177). Fletcher presents a view similar to that of Brown. Accordingly, "the clock hypothesis of relativity theory equates the proper time experienced by a point particle along a timelike curve with the length of that curve as determined by the metric" (Fletcher 2013, 1369). Here, "proper time" means, like with Brown, the time that would be read off by clock along the curve (worldline). Proper time as defined by Minkowski is (1/c of) the length of a timelike curve as determined by the metric, i.e. 1/c $\int ds$. It is not clear if Fletcher considers the clock hypothesis as an independent hypothesis like Brown, or somehow implicit in the theory as defended by Arthur. Almost implicitly, Fletcher takes an ideal clock to be a clock whose time reading is equal to 1/c $\int ds$ (Fletcher 2013, 1371 and 1382). Fletcher considers that in his work he provides "an existence proof for sufficiently ideal light clocks" (Fletcher 2013, 1371).

16 Fletcher's theorem applies to the general case of a curved space-time but here, we will just consider it in relation to the Minkowski space-time. According to Fletcher, his work generalizes, in particular, the previous works by Maudlin (2012) and Gautreau and Anderson (1969) regarding "providing an existence proof for sufficiently ideal light clocks" (Fletcher 2013, 1371). From this point onwards we will follow Fletcher and adopt geometric units, in which c = 1.

In Fletcher's approach, we consider a mirror represented by a $C^{(2)}$ timelike worldline γ. We want to relate the length of a closed segment of this worldline γ[I'] to the elapsed time of a light clock formed by the mirror with worldline γ and a companion mirror whose worldline belongs to a convergent companion family of worldlines $γ_α$, where α is an index that labels the worldlines in the companion family. Each companion worldline, to be such, has a non-zero scalar radius field $r_α$ on γ[I'], which we can see as giving a measure of the distance between the two mirrors (the mirror in γ and the mirror in the companion curve $γ_α$), even if not a privileged one. The requirement that $r_α$ is non-zero "amounts to ensuring that there is a non-zero distance between the mirrors, hence the "photon bouncing" is always well-defined" (Fletcher 2013, 1376). While we take $r_α$ to be non-zero it is imposed the condition that $\lim_{α→∞} r_α = 0$. Each $γ_α$ corresponds to a possible light clock formed by the mirror with worldline γ and the companion mirror with worldline $γ_α$. The condition $\lim_{α→∞} r_α = 0$ implies that we can choose smaller and smaller light clocks. According to Fletcher "having a convergent family of companion curves means that there is always available a sufficiently "small" light clock as determined by the scalar field r" (Fletcher 2013, 1381). This means that we can choose an accuracy $ε_A$ and a maximum variation of regularity $ε_R$ that are both as small as we wish so that we have a light clock that gives a time measure whose difference to the length (Minkowski proper time) is smaller than $ε_A$, and with no more than $ε_R$ variation in regularity. For sufficiently large α we can associate to each $γ_α$ a bounce number $n_α$, which is the number of times the light "bounces" in the companion worldline $γ_α$ (Fletcher 2013, 1377). For each worldline $γ_α$ we can choose a "distance" $d_α$ which lies between the minimum (min $r_α$) and maximum (max $r_α$) values of the "radial distance field" $r_α$ (Fletcher 2013, 1378). One also requires the condition that $\lim_{α→∞}$ max $r_α$ / min $r_α = 1$, which implies that the distance between the mirror in γ and the mirror in each $γ_α$ cannot be too variable. According to Fletcher, one can interpret this condition as requiring that, for sufficiently small light clocks, the order of magnitude of variation in the range of the scalar radius field is small. This makes sense, for if the field's maximum [(max r)] and minimum [(min r)] are not of the same scale, one would expect that the error induced by variation in r is never reduced. (Fletcher 2013, 1381)

Fletcher proves the result that $\lim_{α→∞} 2n_α d_α = |I'|$, where |I'| is the length of the closed timelike wordline segment under consideration (i.e. it is the Minkowski proper time).[17] The physical meaning of the left term of the equality (i.e. $\lim_{α→∞} 2n_α d_α$) is the same as in the simpler case of a light clock in inertial motion in which the two worldlines are parallel with a distance d: "if the light ray completes n round-trips between the mirrors, then the clock has recorded an elapsed time of 2nd" (Fletcher 2013, 1370). That is, $\lim_{α→∞} 2n_α d_α$ is the measured elapsed time gone by the light clock constituted by the mirror with worldline γ and the companion mirror with worldline $γ_α$. Accordingly,

the first limiting equation, concerning the accuracy of the light clock, states that (under the conditions of the theorem) twice the product of this d with the number of observed bounces [which is the elapsed time gone by the light clock along the worldline γ] will approximately equal the [length of the worldline γ (which Minkowski called "proper time")]. (Fletcher 2013, 1382)

As we have seen, when defining the notion of light clocks from the inertial time (of light propagation), we need for consistency to suppose previous to showing it that the light clocks' rates are independent of their past history. If a light clock gives in any "reasonable" worldline (i.e. a $C^{(2)}$ timelike worldline, as Fletcher showed) a time reading (in the limit) equal to the length of the

---

17 In the proof of the theorem, we associate with each $γ_α$ a sequence of arc lengths of segments of γ[I'] "clocked" by the bounces of light in the worldline $γ_α$ (Fletcher 2013, 1377-8). As mentioned, Fletcher also proves in the theorem a result that implies that the light clock can be as regular as one wishes. Accordingly, "the second limiting equation, concerning the regularity of the clock, states that the maximum difference in [the arc lengths] between any two ticks over the course of the clock's run will be small" (Fletcher 2013, 1382).

worldline (i.e. the Minkowski proper time) then it is (in the limit) independent of its past history and we have a self-consistent definition of clock. Fletcher's result shows that the notion of light clocks is consistent – i.e. it shows that, in fact, we can consider the physical systems we call "light clocks" as clocks in the special theory of relativity.[18, 19]

Fletcher result also shows that there is a solution/model of the theory that corresponds to physical systems "lock onto" the arc length parameter of the timelike wordline, and the "ticks" of these physical systems correspond to our notion of inertial time. We do not need to go outside the theory and the notion of inertial time to find a physical system that (approximately) reads off a time whose value along its timelike worldline is equal to the length of the curve (i.e. to the Minkowski proper time). In particular, we do not need the notion of an atomic clock to have a clock that reads off a time along a worldline (approximately) identical to the Minkowski proper time.[20, 21] Neither do

---

18 Thinking in terms of light clocks there is no need for the clock hypothesis or condition. As Fletcher showed, a light clock reads off a time along a timelike workdline that is equal to the length of the worldline without any necessity for the assumption that the rate of the clock does not depend on its acceleration. Neither it is necessary to consider that the clock hypothesis/condition is implicit in the theory as the criterion for an ideal clock. What we can say is that the clock's rate independence of its past history is a requirement of the theory to which the light clock (or another physical system) must comply to actually be considered a clock. In fact, in the theory of relativity, the clock hypothesis/condition seems to be a different statement of the independence of past history requirement. A clock whose rate does not depend on its previous path in space-time or acceleration is a clock whose rate, as determined in the adopted inertial reference frame, depends only on its instantaneous velocity. This is the clock hypothesis/condition. Since the independence of a clock's rate of its past history is a necessary requirement for a physical system to be a clock, a system that does not comply with this requirement cannot be considered a clock in the theory of relativity. We already mentioned that if a clock reads off a time equal to the Minkowski proper time then we may take it to be independent of its past history. We can go the other way around, and considering this requirement we can notice that a physical system that complies with it (i.e. a clock) is a physical system whose time reading is identical to the length of its worldline (called by Minkowski proper time) (see, e.g., Geroch 1972, 8-15). Since a clock to be so must have a rate that is independent of its past history, this implies that the clocks read off a time along a timelike worldline that is equal to the Minkowski proper time. In this way, any (relativistic) clock to be so must be a clock that is ideal in Arthur's sense of reading off a time that is equal to the Minkowski proper time (and evidently in Brown's sense of reading inertial time while in inertial motion). We do not need to consider that there is a clock hypothesis/condition that gives the criterion for an ideal clock since we do not need a notion of an ideal clock as defined by Arthur (or Brown).

19 We departure from the view that in the theory of relativity there is a bifurcation of the notion of time in terms of the inertial time (or coordinate time) and the proper time as some authors defend (see, e.g., Arthur 2008; Savitt 2011). According to Arthur, in the theory of relativity there is a bifurcation in the conception of time, existing two separate time concepts (the coordinate time and the proper time) with different roles in the theory (Arthur 2007; Arthur 2008). Here we do not regard "proper time" as a separate time concept that can be seen as an independent physical quantity. The total time read off by a light clock along a non-inertial worldline results from the (inertial) propagation of light; it is "built" from inertial time. Regarding the issue of the different roles of proper time and coordinate time, we do not address this issue here.

20 In Bacelar Valente (2016) it is argued that "it is difficult to maintain a complete independence of the Minkowski proper time as a physical quantity from the experimental results about the [empirical] proper time of atomic clocks since these are in part taken into account in the development of the theory" (Bacelar Valente 2016, 201). This follows from the fact that "[atomic] clocks enter the theory's structure in the articulation of the time coordinate of an inertial reference frame and almost as a hidden assumption – the boostability assumption –, necessary in Einstein's view to derive the Lorentz transformations. As such, the existence of [atomic] clocks is not something that is predicted by the theory but a physical input assumption." (Bacelar Valente 2016, 206). According to this paper, "Minkowski's contribution made explicit [that] the theory is able to 'find a place' in its theoretical structure to a concept of a clock that corresponds to the experimental notion of [atomic] clocks. With the Minkowski proper time, we can fit into the theory the empirical proper time of [atomic] clocks that is (at least in part) implicit in its construction." (Bacelar Valente 2016, 206). The atomic time would "enter" the theory as the conceptual clock taken to be independent of its past history and also in relation to the so-called boostability assumption, which can be seen as a particular case of the "independence from past history" assumption (a similar point might have been made in relation to the invariance of ds, "strengthening" this line of argument). It would be in part implicit in the formulation of the Minkowski proper time and would "substantiate" a notion of a clock that reads off a time equal to the (Minkowski) proper time. Here, it is presented the view that the theory can be formulated just in terms of the inertial time. The notion of light clock developed from the inertial time (of light propagation) "substantiates" a notion of a clock that gives an inertial time that along a non-inertial worldline is equal to the length of the curve (i.e. the Minkowski proper time). There is no need for atomic clocks even in the limited sense of being clocks that read off a proper time; the theory already has "solutions" corresponding to identical physical systems (which we call light clocks) that already read off a proper time.

21 In Bacelar Valente (2016) it is defended that the "empirical proper time" of atomic clocks has a relevant role in the we need to resort to the notion of atomic time or atomic clock elsewhere. As we have seen we can formulate the notion of inertial time in the theory of relativity – in terms of the propagation of light, and a (self-consistent) derived notion of light clocks. Contrary to Einstein's approach we can develop the theory solely in terms of the inertial time and the derived notion of light clocks.

4. Conclusions

In this paper, we made the case that the notion of inertial time as expressed in the (inertial) propagation of light is the only temporal notion necessary in the theory of relativity. In particular, there is no need for the notion of atomic time or atomic clock in the theory. From the inertial time of light, we derive the notion of (identical) light clocks. This notion of clocks can be seen as being implicit in the theory, since it is a derived notion from the propagation of light/inertial time. These clocks when synchronized enable to measure the coordinate time. However, we need to show that light clocks have a rate that is independent of their past history to argue that they provide a self-consistent notion of a clock, i.e. that there is, in fact, a notion of a clock that is a "solution" of the theory. Otherwise, we would need to resort to a notion external to the theory – e.g. the atomic clock.

    Taking into account that the theory seems to imply that the time measured along a timelike worldline is equal to the length of the worldline (called by Minkowski "proper time"), it results that if there are clocks that read off a time along a (non-inertial) timelike worldline equal to the length of the worldline, these clocks are independence from their past history. The light clocks turn out to be clocks that read off a time that approximates to an arbitrary degree the Minkowski proper time. In this way, they have a rate that is approximately (to an arbitrary degree) independent of their past history. This implies that the adoption of light clocks, which we consider to be solutions/models of the theory, is consistent.

---

theory. This view is developed by accepting as a premise Einstein's view that the concept of a clock must be considered an independent self-sufficient concept, since, according to him, the theory does not have solutions corresponding to physical systems to which we might ascribe the role of clocks. In this work, we reject Einstein's view. This leads to the view that atomic clocks are not necessary for the foundations of the theory. We can formulate the theory entirely in terms of the inertial time. We can consider "alternative" formulations in which atomic time has a role (see, e.g., Bacelar Valente 2017), but this is not a necessity, it is simply a possibility.